\documentclass[prl,twocolumn,showpacs,preprintnumbers,amsmath,amssymb,floats,citeautoscript]{revtex4-1}
\usepackage{graphicx}
\usepackage{dcolumn}
\usepackage{bm}
\usepackage{txfonts}
\usepackage{multirow}
\usepackage{here}

\begin{document} 

\title{Enhanced Seebeck coefficient by a filling-induced Lifshitz transition in K$_{x}$RhO$_{2}$}

\author{Naoko~Ito}
\email{6217603@ed.tus.ac.jp}
\author{Mayu~Ishii}
\author{Ryuji~Okazaki}
\email{okazaki@rs.tus.ac.jp}

\affiliation{Department of Physics, Faculty of Science and Technology, Tokyo University of Science, Noda 278-8510, Japan}

\begin{abstract}

We have systematically measured the transport properties in the layered rhodium oxide 
K$_{x}$RhO$_{2}$ single crystals ($0.5\lesssim x \lesssim 0.67$), 
which is isostructural to the thermoelectric oxide Na$_{x}$CoO$_{2}$. 
We find that below $x = 0.64$ the Seebeck coefficient is anomalously enhanced at low temperatures
with increasing $x$, 
while it is proportional to the temperature like a conventional metal above $x=0.65$, suggesting an existence of a critical content $x^{*} \simeq 0.65$.
For the origin of this anomalous behavior, we discuss a filling-induced Lifshitz transition, 
which is characterized by a sudden topological change in the cylindrical hole Fermi surfaces at the critical content $x^*$.

\end{abstract}

\maketitle

The layered cobalt oxide Na$_{x}$CoO$_{2}$ has attracted a great deal of attention for 
its rich physical properties since the discovery of the large Seebeck coefficient \cite{Terasaki1997}. 
This compound consists of alternately-stacked CoO$_2$ and Na layers, and 
the conductive CoO$_2$ layer is formed by edge-shared CoO$_{6}$ octahedra, resulting in a two-dimensional (2D) triangular lattice of Co ions.
The uniqueness of this compound is the complex electronic properties drastically changing
with Na content $x$ \cite{Foo2004}.
For rich Na content region, the large Seebeck coefficient is observed with metallic resistivity \cite{Motohashi2001,Lee2006}. 
The magnetic property varies from Pauli-paramagnetic to Curie-Weiss (CW) behavior across a critical Na content 
$x^*_{\rm Na} \simeq 0.62$ 
\cite{Yokoi2005,Yoshizumi2007,Okamoto2010,Lang2008}. 
A spin-density-wave and a charge-ordered insulating state have also been suggested for $0.75\leq x$ and at $x\sim$ 0.5, respectively 
\cite{Motohashi2003,Sales2004,Huang2004,Sugiyama2004,Ning2008}. 
Moreover, hydration induces unconventional superconductivity at around $x=0.3$ \cite{Takada2003,Jin2003,Mazin2005,Yang2005PRB,Ogata2007}.

The crucial points for understanding various emergent phenomena in Na$_{x}$CoO$_{2}$ are 
the characteristic electronic band structure and the topology of the Fermi surface.
In this compound,
the $t_{{\rm 2}g}$ bands from Co 3$d$ orbitals are composed of an $a_{{\rm1}g}$ band and doubly degenerate $e_{g}'$ bands, 
and the Fermi level lies across the $a_{1g}$ band to form
a cylindrical Fermi surface reflecting the quasi-2D crystal structure \cite{Singh2000,Korshunov2007,Kuroki2007PRL,Hasan2004,Yang2005,Yang2004,Qian2006,Geck2007,Arakane2011}.
Now the shape of the $a_{1g}$ band is peculiar, 
including a somewhat flat portion 
around a local minimum at the $\Gamma$ point,
which is 
called pudding-mold band shape \cite{Kuroki2007JPSJ}.
With increasing Na content, the Fermi level rises and then touches the local minimum to create 
a small electron pocket as is resolved by the angle resolved photoemission spectroscopy (ARPES) \cite{Arakane2011}.
Detailed thermodynamic study has revealed that 
this topological change in the Fermi surface, known as the Lifshitz transition \cite{Lifshitz1960}, 
occurs at the critical content $x^*_{\rm Na} \simeq 0.62$ \cite{Okamoto2010}.
The Fermi level then locates around the flat region for $x>x^*_{\rm Na}$, 
resulting in the large Seebeck coefficient due to a significant difference in the velocities of electrons and holes \cite{Kuroki2007JPSJ}.
The flat band with large density of states may also contribute to the CW-like magnetic behavior observed above $x^*_{\rm Na}$.

The issue to be addressed is whether the richness of these intriguing properties universally
emerges in the related systems.
K$_x$CoO$_2$ ($x \sim 0.5$)
shows two phase transitions at low temperatures \cite{Yokoi2008,Watanabe2006},
reminiscent of Na$_{0.5}$CoO$_2$, and
similar electronic structures in these materials are suggested both experimentally and 
theoretically \cite{Qian2006,Lee2007}.
The $4d$ system Na$_x$RhO$_2$ is also isostructural to Na$_{x}$CoO$_2$ \cite{Varela2005,Krockenberger2007}
and exhibits a metal-insulator transition
which is explained by an Ioffe-Regel criterion \cite{Zhang2018},  
but the detailed electronic properties are still unexplored.

The isostructural K$_{x}$RhO$_{2}$ 
shows moderately large Seebeck coefficient with half the magnitude of Na$_{x}$CoO$_{2}$ \cite{Yubuta2009,Shibasaki2010,Yao2012}. 
The optical study suggests that the bandwidth of K$_{x}$RhO$_{2}$ is twice broader compared with Na$_{x}$CoO$_{2}$ 
owing to a difference of the orbital sizes between Co 3$d$ and Rh 4$d$ electrons \cite{Okazaki2011}.
Recent ARPES measurement has 
clarified that the band shape of K$_{x}$RhO$_{2}$ is closely similar to 
that of Na$_{x}$CoO$_{2}$ \cite{Chen2017}, 
indicating a possible emergence of rich electronic phase diagram.
Furthermore, interesting phenomena,  including 
a significant enhancement of thermoelectric efficiency by hydration \cite{Saeed2012} and
a realization of topological quantum Hall effect \cite{Zhou2016},
have been theoretically predicted in K$_{x}$RhO$_{2}$.
However, the physical properties of K$_{x}$RhO$_{2}$ have been little known 
due to the difficulty of the synthesis of
well controlled K content, single-phase samples \cite{Zhang2013,Zhang2016}. 

In this paper, we present
the potassium composition dependence of the transport properties in K$_{x}$RhO$_{2}$ for $0.5\lesssim x\lesssim0.67$
using  pure single-crystalline samples, which 
are systematically prepared by a self-flux method and a K de-intercalation process.
We find that the Seebeck coefficient is highly sensitive to the K content $x$ and 
enhanced only below a critical content $x^{*}\simeq 0.65$. 
This result is attributed to a filling-induced Lifshitz transition at $x^*$,
at which the topology of the cylindrical hole surface dramatically changes.
Moreover, near $x=0.5$, distinct anomaly is observed at $T\simeq75$~K,
implying that this rhodate also offers a fascinating platform for various electronic phases to be investigated.

Single-crystalline samples of K$_{x}$RhO$_{2}$ were grown by a self-flux method \cite{Shibasaki2010,Yubuta2009}. 
A mixture of K$_{2}$CO$_{3}$ (99.999\%) and Rh$_{2}$O$_{3}$ (99.9\%) with a molar ratio of $50:1$ was put in an alumina crucible and 
kept at 1373~K for 1~h, and then cooled down to 1123~K with a rate of 5~K/h. 
After washing with distilled water, 
we obtained thin hexagonal crystals 
as shown in the inset of Fig. 1(a). 
X-ray diffraction (XRD) measurements were performed by an X-ray diffractometer (Rigaku UltimaIV) with Cu K$\alpha$ radiation in a $\theta$-2$\theta$ scan mode. 
In the case of the single crystalline samples, the scattering vector was normal to the surface of the sample. 
The K content of as-grown samples is $x\sim0.67$. 
To get the lower potassium content samples, 
we dipped the samples in I$_{2}$-acetonitrile solutions \cite{Zhang2016}. 
The potassium content was controlled by both the concentration of the solution and the length of time to dip. 
This process was conducted at room temperature. 
Subsequently, we annealed these dipped samples at 473 K for about 2 days in air to obtain single phase samples. 
To elucidate the relation between the $c$-axis lattice parameter and potassium content, 
we measured the K content $x$ using an electron probe microanalyzer (JEOL JXA-8100) 
after we determined the $c$-axis lattice parameter by XRD measurement. 
The in-plane resistivity and the Seebeck coefficient were simultaneously measured by using a conventional dc four-probe method and a steady-state method. 
The thermoelectric voltage contribution from the wire leads was subtracted.

\begin{figure}[tb]
\includegraphics[width=1\linewidth]{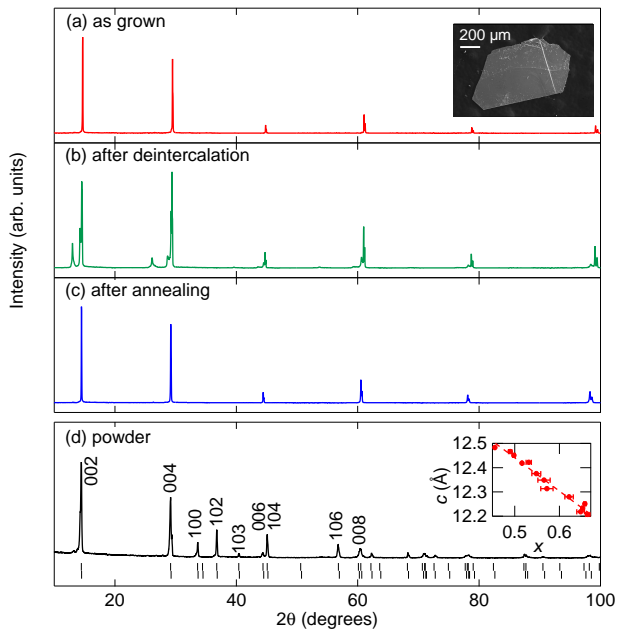}
\caption{(Color online).
(a-c) The XRD pattern of the single crystal: (a) as grown, (b) after de-intercalation, and (c) after annealing. 
The inset of (a) shows the SEM image of as-grown K$_{x}$RhO$_{2}$ crystal.
(d) The powder XRD pattern. 
Bars indicate the expected Bragg peak positions for $\gamma$-type structure K$_{x}$RhO$_{2}$. 
The inset shows the relation between the K content $x$ and $c$-axis lattice parameter $c$. 
The dashed line represents a linear fitting result. 
}
\end{figure}

We first show the XRD patterns of a single crystal in each process of the K de-intercalation.
The XRD patterns of  as-grown and after-de-intercalation crystals are depicted in Figs. 1(a) and 1(b), respectively.
As reported by Zhang $et~al$ \cite{Zhang2013,Zhang2016},
several sets of the peaks are observed in after-de-intercalation crystals.
On the other hand, after annealing, we find that a single-phase pattern is recovered as shown in Fig. 1(c).
In the isostructural Na$_{x}$CoO$_{2}$, it is well known that 
the low Na-content sample is easily hydrated 
and a heating process removes H$_{2}$O molecules from the hydrated sample \cite{Chen2004}. 
Thus the present result indicates that a hydrated phase is also formed in K$_{x}$RhO$_{2}$
after de-intercalation but
can be recovered to the non-hydrate phase by the successive annealing.
The existence of the hydrated K$_{x}$RhO$_{2}$ is also suggested by
an ion exchange experiment \cite{Mendiboure1987}.
Note that no superconductivity is observed in the hydrated K$_{x}$RhO$_{2}$ at present.

The powder XRD pattern of non-hydrate sample for $x= 0.651$ is represented in Fig. 1(d). 
Note that the measured powder is not ground and just collected as small single crystals with the size of $\sim 50$ $\mu$m,
because the grinding induces the hydration significantly.
As seen in Fig. 1(d), all the peaks are indexed using the $\gamma$-Na$_x$CoO$_2$-type structure (space group $P6_{3}/mmc$)
with the lattice parameters of $a = 3.076(2)~\AA$ and $c = 12.233(5)~\AA$.
In the earlier reports, 
the $c$-axis lattice parameter is found to be 13.6 $\AA$ \cite{Yubuta2009,Yao2012},
much larger than the present result.
This discrepancy indicates that
the powder XRD patterns in the earlier reports 
refer to the hydrated phase, 
as is also pointed out by means of the optimization of the lattice constants \cite{Saeed2012}. 
On the other hand, the space group is preserved
among the hydrated and non-hydrate samples, 
similar to the case of Na$_{x}$CoO$_{2}$ \cite{Takada2003}. 

\begin{figure}[t]
\includegraphics[width=8cm]{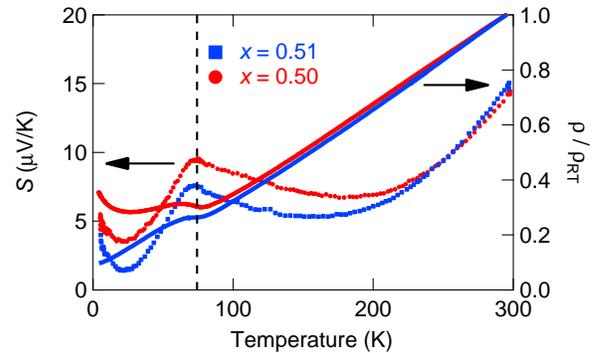}
\caption{(Color online).
Temperature dependence of the Seebeck coefficient (left axis) and 
the resistivity normalized by the room-temperature resistivity $\rho_{\rm RT}$ (right axis) of 
K$_{x}$RhO$_{2}$ for $x=0.50$ and 0.51. 
The dashed line represents the temperature at which the distinct anomaly is observed. 
}
\end{figure}

We examine the relation between the $c$-axis lattice parameter and 
the K content $x$ using non-hydrate crystals 
as shown in the inset of Fig. 1(d).
The $c$-axis lattice parameter decreases almost linearly with increasing K content $x$. 
In this system, when the K content increases, 
the Coulomb force holding the RhO$_2$ layers is enhanced to decrease the $c$-axis lattice parameter,
which is also seen in Na$_x$CoO$_2$ \cite{Foo2004}.
The present data are fitted using a linear function $c=c_0+c_1x$
with $c_0=13.13(4)$ $\AA$ and $c_1=-1.37(7)$ $\AA$
as shown by the dashed line.
The K content of crystals for the transport measurements is calculated 
from the $c$-axis lattice parameter using this relation in a similar manner to that of Na$_x$CoO$_2$ \cite{Yokoi2005}.
The $x$ values have a common error which is estimated at up to 0.04 due to the large error of $c_0$ and $c_1$.
However, we can obtain relative amount of K content accurately by the comparison of the $c$-axis lattice parameter.
From the K content, the hole concentration is approximately determined as $1-x$ holes/Rh \cite{Chen2017}.

We then discuss the transport properties of K$_{x}$RhO$_{2}$.
Figure 2 shows the temperature dependence of the Seebeck coefficient and the resistivity for $x=0.50$ and 0.51. 
The resistivity is normalized by the room-temperature resistivity $\rho_{\rm RT}\sim 1$ m$\Omega$cm,
which has a large error because the sample thickness is too small to be measured correctly.
We also note that the value of $\rho_{\rm RT}$ does not change significantly for all the crystals in the present study.
A distinct anomaly is observed in both the Seebeck coefficient and resistivity at around 75 K, 
as marked by the dashed line. 
An origin of this anomaly is not clear at present.
In Na$_{x}$CoO$_{2}$, the two distinct phase transitions were observed, 
and the ground state is a charge-ordered insulator at around $x=0.5$ \cite{Foo2004,Yokoi2005}. 
Moreover, an interesting topological state has been suggested theoretically for $x=0.5$ \cite{Zhou2016}.

\begin{figure}[t]
\includegraphics[width=9cm]{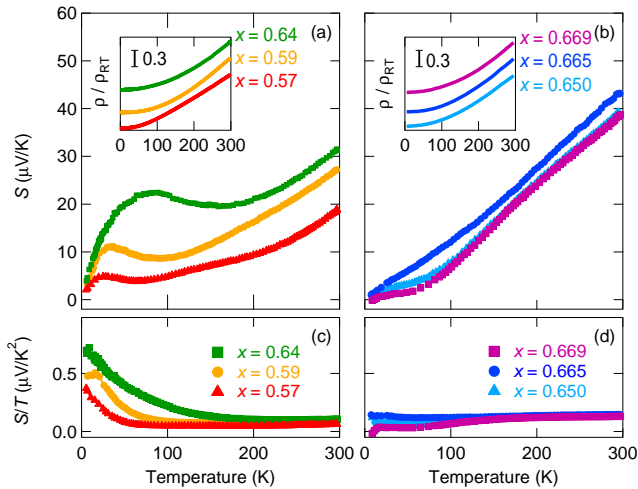}
\caption{(Color online).
The temperature dependence of (a-b) the Seebeck coefficient and 
(c-d) the Seebeck coefficient divided by temperature, $S/T$. 
The left panels show the data below $x = 0.64$ 
and the right panels show above $x= 0.65$.
The insets show the normalized resistivity as a function of temperature.
}
\end{figure}

We next show the temperature dependence of the transport properties for $0.57\lesssim x\lesssim0.67$ in Figs. 3(a) and (b). 
Around room temperature, the Seebeck coefficient almost linearly decreases with decreasing temperature 
for all the samples. 
On the other hand, 
the temperature dependence does not systematically vary with the K content. 
For $x\gtrsim0.65$, the Seebeck coefficient is roughly proportional to temperature as shown in Fig. 3(b). 
In contrast, as shown in Fig. 3(a), the temperature dependence for $0.57\lesssim x\lesssim0.64$ 
is qualitatively 
different from that for $x\gtrsim0.65$:
At low temperatures, the Seebeck coefficient is enhanced with a broad peak structure.
Note that the phonon drag effect is unlikely since we have observed similar temperature variation even in the 
polycrystalline samples (not shown).

\begin{figure}[t]
\includegraphics[width=8.5cm]{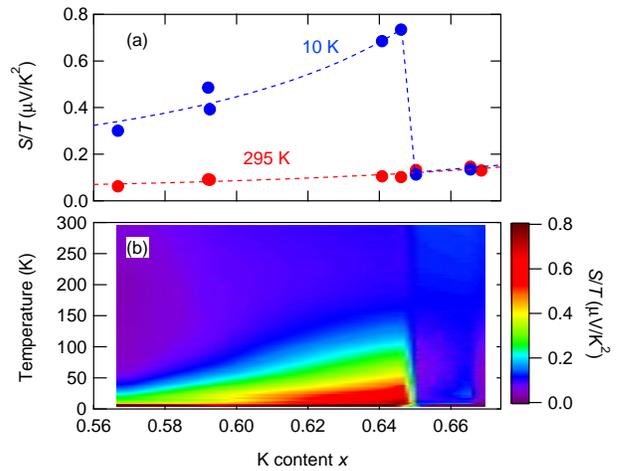}
\caption{(Color online).
(a) The $x$ dependence of $S/T$ of K$_{x}$RhO$_{2}$ at $T=10$~K (blue) 
and $295$~K (red). 
The dashed lines are guides for eyes. (b) Contour plot of $S/T$.
}
\end{figure}

The anomaly is
clearly demonstrated in the Seebeck coefficient divided by temperature, $S/T$, shown in Figs. 3(c) and (d). 
For $0.57\lesssim x\lesssim0.64$, 
$S/T$ gradually increases with lowering temperature.
It also increases with increasing $x$ 
but the behavior drastically changes in the narrow range of $0.64<x<0.65$
into almost no temperature dependence, as expected in a conventional metal,
suggesting that there is a critical content $x^{*}$ in this narrow range. 
Figure 4(a) represents $S/T$ as a function of $x$ at constant temperatures
and the contour plot is depicted in Fig. 4(b).
At low temperature, 
$S/T$ shows a divergent behavior at $x^*$ only from the lower $x$ side.
This is distinct from 
the divergence of $S/T$ from both sides of the critical point
of the Landau-type phase transition
associated with broken symmetry \cite{Sakai2016}.
Also, no structural transition is observed at  $x^*$.
It is a key point that the enhancement is observed for only low temperature,
while $S/T$ continuously increases with $x$ near room temperature,
which is readily explained by the change in the carrier concentration.

Now we discuss the origin of the enhancement of the Seebeck coefficient.
According to the observed band structure of K$_{0.62}$RhO$_{2}$, 
the Fermi level lies across the $a_{1g}$ band \cite{Chen2017}. 
The Fermi level is elevated as $x$ increases, 
and as is the case in Na$_{x}$CoO$_{2}$, 
the topological change in the Fermi surface would also occur at a certain content in K$_{x}$RhO$_{2}$. 
Note that
the present critical content $x^*$ is close to $x^*_{\rm Na}\simeq0.62$, 
at which the Lifshitz transition occurs in Na$_{x}$CoO$_{2}$ \cite{Okamoto2010}.
Moreover, near the Lifshitz transition,
the low-temperature Seebeck coefficient is predicted to increase steeply 
when the Fermi level approaches a critical point 
from only one side, the one for
which the number of Fermi surfaces is larger, 
owing to the divergence in the energy derivative of the density of states
\cite{Varlamov1989,Abrikosovbook},
as is also explored in several alloys experimentally \cite{Egorov1982,Egorov1984}.
Thus the observed enhancement indicates the existence of a Lifshitz transition in K$_{x}$RhO$_{2}$. 
We also note that the resistivity has no singularity around the Lifshitz transition \cite{Varlamov1989}.
In the insets of Figs. 3(a) and (b),
we show the temperature dependence of the normalized resistivity, which exhibits no significant change across $x^*$, 
consistent with above picture.

\begin{figure}[t]
\includegraphics[width=8cm]{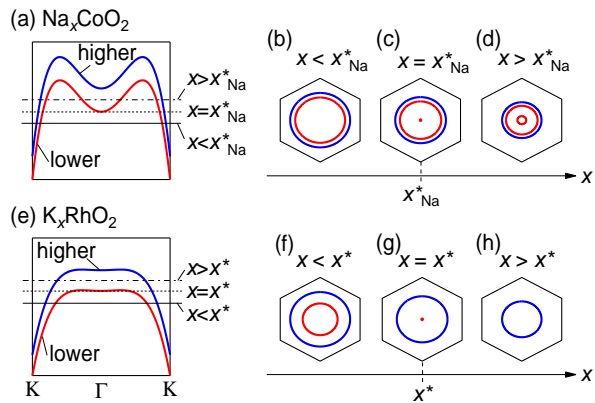}
\caption{(Color online).
Schematic pictures of (a) the band structure along K-$\Gamma$-K line and
(b-d) the Fermi surface as a function of the Na content $x$ of Na$_{x}$CoO$_{2}$.
(e-h) Corresponding band structure and the Fermi surface of K$_{x}$RhO$_{2}$.
The red and blue curves represent the lower and higher bands respectively, in common with the band structure and the Fermi surface. 
Near the $\Gamma$ point, the band is more flat in K$_{x}$RhO$_{2}$ compared with that of Na$_{x}$CoO$_{2}$. 
As $x$ increases, the electron pocket suddenly appears at $x=x^{*}_{\rm Na}$ in Na$_{x}$CoO$_{2}$ 
while the inner hole cylinder gradually diminishes and then disappears at $x=x^{*}$ in K$_{x}$RhO$_{2}$.
}
\end{figure}

We examine the detailed mechanism of the Lifshitz transition.
In Fig. 5, we show the schematic band structure and the Fermi surface of Na$_{x}$CoO$_{2}$ and K$_{x}$RhO$_{2}$.
Since the $\gamma$-type structure has two Co/Rh ions in the unit cell, 
we consider two $a_{1g}$ bands labelled as higher and lower bands in Figs. 5(a) and 5(e) \cite{Singh2000},
although these are almost degenerate and
difficult to be resolved separately by ARPES measurements.
In Na$_{x}$CoO$_{2}$, there are two hole cylinders for $x<x^{*}_{\rm Na}$ as shown in Fig. 5(b), and 
with increasing $x$, the Fermi level rises and then touches the local minimum of the lower band at $x=x^{*}_{\rm Na}$ [Fig. 5(c)] to create 
a small electron pocket at the $\Gamma$ point for $x>x^{*}_{\rm Na}$ [Fig. 5(d)].
In this case, 
the low-temperature Seebeck coefficient would be critically enhanced owing to 
the disappearance of the electron pocket
only when $x$ approaches $x^{*}_{\rm Na}$ from the high $x$ side.
Indeed, negatively enhanced Seebeck coefficient is seen for $x>x^{*}_{\rm Na}$,
although the experiments have been done in polycrystalline samples \cite{Okamoto2010}.

On the other hand, the enhancement of the Seebeck coefficient is observed for $x<x^{*}$ in K$_{x}$RhO$_{2}$, 
indicating that the topological change in the Fermi surfaces differs from that of Na$_{x}$CoO$_{2}$.
According to recent ARPES measurement \cite{Chen2017}, 
the band structure of K$_{x}$RhO$_{2}$ is closely similar to that of Na$_{x}$CoO$_{2}$,
and thus
we consider a local minimum structure at the $\Gamma$ point as well.
In Na$_{x}$CoO$_{2}$, this minimum structure is induced by a third nearest hopping $t_{3}$, 
and the local minimum sinks more deeply 
as the ratio $|t_{3}/t_1|$ becomes larger ($t_1$ being the first nearest hopping) \cite{Kuroki2007JPSJ}. 
Now, the 4$d$ orbital is broader than 3$d$ one, but the ionic radius of Rh is also larger than Co,
resulting in an accidental cancelation of the effect of the broader $4d$ orbital in some materials \cite{Okada2005}. 
In K$_{x}$RhO$_{2}$, the broader bandwidth is indeed observed \cite{Okazaki2011,Chen2017}, 
but $|t_{3}/t_1|$ may be smaller than that of Na$_{x}$CoO$_{2}$ 
because oxygen $2p$ contribution in $t_{3}$ would be small due to the larger Rh-Rh distance.
In such a case, 
the local minimum structure of the band top would be shallow in K$_{x}$RhO$_{2}$ as illustrated in Fig. 5(e),
also implied by the DFT calculations \cite{Chen2017}.

This scenario is schematically depicted in Figs. 5(e-h).
In K$_{x}$RhO$_{2}$, the shallow local minimum structure may be undetectable because of thermal fluctuations even at low temperatures.
For increasing $x$, 
the Fermi level is elevated and then the lower $a_{1{\rm g}}$ band is fully occupied at $x^{*}$. 
In this case, the inner hole cylinder gradually diminishes and then completely disappears for $x>x^{*}$,
which is also supported by a highly 2D character of the Fermi surface \cite{Chen2017}.
Thus, in K$_{x}$RhO$_{2}$, 
the low-temperature enhancement of $S/T$ occurs only for $x<x^{*}$,
where the number of hole cylinders is larger than that for $x>x^{*}$.
It should be noted that
the disappearance of the electron pocket in Na$_{x}$CoO$_{2}$ contributes to the Seebeck coefficient negatively but
the disappearance of the hole cylinder in K$_{x}$RhO$_{2}$ contributes positively,
enhancing the total Seebeck coefficient that also includes the contribution from the outer hole cylinder.
Therefore, our results suggest the Lifshitz transition as a fundamental means to increase the Seebeck coefficient at low temperatures in this system, 
in addition to the previously reported pudding-mold band structure \cite{Kuroki2007JPSJ} 
and the large entropy flow of the $d$ electrons \cite{Koshibae2000}.

To summarize, we have synthesized single crystal K$_{x}$RhO$_{2}$ for $0.50\lesssim x\lesssim0.67$ and 
find that, while the Seebeck coefficient monotonically increases with increasing $x$ near room temperature,
it exhibits an enhancement at low temperatures only below $x^*\simeq0.65$.
As an origin, we propose a Lifshitz transition associated with disappearance of the inner hole cylinder due to the flat band structure around the $\Gamma$ point.
Moreover, we have discovered a distinct phase transition for $x\sim$ 0.5, 
suggesting that 
K$_{x}$RhO$_{2}$ also offers a fascinating platform for various electronic phases,
similar to Na$_{x}$CoO$_{2}$.

\begin{acknowledgments}
We thank K. Fujimoto, H. Yaguchi, T. Yamanaka for discussion and 
D. Kabasawa, W. Takagi for experimental supports.
This work was supported by JSPS KAKENHI Grants No. JP17H06136, No. JP18K03503, and No. JP18K13504.

\end{acknowledgments}

\end{document}